\documentclass[12pt,preprint]{aastex}
\usepackage{rotate}
\usepackage{color}
\usepackage{graphicx}
\shorttitle{No evidence for adiabatic walking}
\shortauthors{Melikidze et al.}
\begin{document}

\def\ni{\noindent}
\def\be{\begin{equation}}
\def\ee{\end{equation}}
\def\lesssim{\raisebox{-0.3ex}{\mbox{$\stackrel{<}{_\sim} \,$}}}
\def\gtrsim{\raisebox{-0.3ex}{\mbox{$\stackrel{>}{_\sim} \,$}}}

\title{On the adiabatic walking of plasma waves in a pulsar magnetosphere}

\author{George I. Melikidze\altaffilmark{1,}\altaffilmark{2}}
\author{Dipanjan Mitra\altaffilmark{3}}
\author{Janusz Gil\altaffilmark{1}}
\affil{\altaffilmark{1}Kepler Institute of Astronomy, University of Zielona Gora, Lubuska 2, 65-265
Zielona G\'ora, Poland} \affil{\altaffilmark{2}Abastumani Astrophysical Observatory, Ilia State
University, 3-5 Cholokashvili Ave., Tbilisi, 0160, Georgia}
\affil{\altaffilmark{3}National Centre for Radio Astrophysics Ganeshkhind, Pune 411 007 India}
\email{gogi@astro.ia.uz.zgora.pl \\ dmitra@ncra.tifr.res.in \\ jag@astro.ia.uz.zgora.pl}

\begin{abstract} The pulsar radio emission is generated in the near magnetosphere of the neutron
star and it has to propagate through the rest of it to emerge into the interstellar medium. An
important issue is whether this propagation affects the planes of polarization of the generated
radiation. Observationally, there is a sufficient evidence that the emerging radiation is polarized
parallel or perpendicular to the magnetic field line planes that should be associated with the
ordinary O and extraordinary X plasma modes respectively, excited by some radiative process. This
strongly suggests that the excited X- and O-modes are not affected by the so-called Adiabatic
Walking that causes a slow rotation of polarization vectors. In this paper we demonstrate that the
conditions for the Adiabatic Walking are not fulfilled within the soliton model of pulsar radio
emission, in which the coherent curvature radiation occurs at frequencies much lower than the
characteristic plasma frequency. The X-mode propagates freely and observationally represents the
primary polarization mode. The O-mode has difficulties in escaping from the pulsar plasma,
nonetheless it is sporadically observed as a weaker secondary polarization mode. We discuss a
possible scenario under which the O-mode can also escape from the plasma and reach an observer.
\end{abstract}


\section{Introduction}
The brightness temperature (exceeding $10^{12}$ K by many orders of magnitude) deduced from the
observed radio flux densities strongly implies that the pulsar radiation must be emitted
coherently. Generally, the coherent pulsar radio emission can be generated by means of either a
maser or a coherent curvature mechanism \citep[e.g. ][]{gz75,rs75,mp80,mp84,kmm91,mgp00}. There is
general agreement that this radiation is emitted in strongly magnetized electron-positron plasma
well inside the light cylinder. Many observational constraints on emission altitudes imply that the
emitted radiation detaches from the ambient plasma at altitude $r_d$ less than 10\% of the light
cylinder radius\footnote{Let us note that this may not be true for the millisecond pulsars.}
$R_{LC}=Pc/2\pi$
\citep[e.g.][]{cor78,bcw91,r93,kg97,md99,gg03,ml04,kmgkga09}.
The emission is clearly emitted at a lower altitude $r_{em}$ and it must propagate through the
magnetosphere to reach an observer, that is $r_{em}\le r_d$. The plasma properties in the region
between $r_{em}$ and $r_d$ can, in principle, influence the nature of waves, and in Section
\ref{secAWC} we will discuss propagation effects, especially those that concern the polarization
state.

Once the waves are generated in the emission region ($r\sim r_{em}$), then in the propagation
region ($r_{em}<r<r_{d}$) they naturally split into the ordinary O-mode and extraordinary X-mode,
which correspond to the normal modes of wave propagation in the strongly magnetized plasma
\citep[see e.g.][]{ab86,lmmp86}. The ordinary waves are polarized in the plane of the wave vector
${\bf k}$ and the local magnetic field ${\bf B}$ and their electric field has a component along
both ${\bf k}$ and ${\bf B}$. Therefore, they interact strongly with plasma particles and thus
encounter difficulty in escaping from the magnetosphere. On the other hand, the extraordinary waves
are linearly polarized perpendicularly to ${\bf k}$ and ${\bf B}$. As a result the X-mode can
propagate through the magnetospheric plasma almost as in vacuum.

It is important to realize that there is also observational evidence confirming the extraordinary
mode to be dominant in the pulsar radiation. \citet{lcc01}, based on the X-ray image of the Vela
pulsar wind nebula were able to model the pulsars rotation axis as it is projected in the sky. The
X-ray image showed two X-ray arcs that were interpreted to be a result of outflowing relativistic
particles from two diametrically opposite pulsar beams interacting with the environment around the
pulsar. The bisection of the two arcs gave the direction of the rotation axis projected on the sky
plane. Further they used the radio polarization property of the Vela pulsar where the linear
polarization position angle traverse is well represented by the so called rotating vector model
(RVM). In this model the polarization position angle (PPA) of the linear polarization executes a
s-shaped track across the pulses. This was interpreted by \citet{rc69} as electric field vector
associated with the range of open dipolar magnetic field planes intersected by the observer's line
of sight.  The steepest gradient (SG) point in the PPA track is usually considered as the fiducial
point located in the plane containing the magnetic dipole and the rotation axis (at least in slow
pulsars).
\citet{lcc01} used the value of the absolute PPA at the SG point (PA$_{\circ}$), and found its
direction to be perpendicular to the rotation axis, which means that the electric vector emanating
out of the Vela pulsar is orthogonal to the dipolar magnetic field line planes. Hence it was the
extraordinary (X) mode \citep{lcc01}. This is a very significant observational result, as for the
first time it was possible to verify that the electric field vectors emerging from the Vela pulsar
magnetosphere are perpendicular to the dipolar magnetic field line planes. The Vela X-ray
observations also showed that the proper motion direction (PM$_{v}$) of the pulsar is aligned with
the rotation axis.

An X-ray pulsar wind nebula is not observed for majority of the pulsars, and hence the direction of
the rotation axis on the plane of the sky cannot be determined directly. However, based on several
careful recent studies by \citet{john05}, \citet{r07}, and \citet{nkcj12,nskkj13}, a bimodal
distribution of $\Psi$=(PM$_{v}$ - PA$_{\circ}$) centered around 0$^{\circ}$ and 90$^{\circ}$ has
been observed. \citet{mrg07} used this distribution, along with the assumption that the pulsar spin
axis is aligned with the proper motion direction to establish the direction of the emerging wave at
PA$_{\circ}$ with respect to the magnetic field line planes.

It is instructive to follow and generalize their arguments. The proper motion measurement of many
pulsars can give the projected orientation of the rotation axis with respect to the celestial
north.  We can also independently measure the absolute position angle PA$_{\circ}$ (corrected for
all instrumental and propagation effects) at the fiducial point (SG point and/or profile midpoint).
Then we can find the difference $\Psi$ between these angles, and amazingly $\Psi$ shows a bimodal
distribution centered around 0$^{\circ}$ and 90$^{\circ}$. Let us ask the question how this can
happen? It is possible only if the spin axis is either parallel or perpendicular to the direction
of the proper motion and at the same time the PA$_{\circ}$ coincides either with the fiducial plane
or the plane perpendicular to it. Let us assume for the moment that PA$_{\circ}$ is at some
arbitrary angle with respect to the fiducial plane. Then we have to find such a direction for the
velocity vector which is either perpendicular or parallel to the polarization vector at the
fiducial point. Then this needs to be extended to all pulsars in the sample, in which every pulsar
could have different PM$_{v}$ and PA$_{\circ}$. Such fine tuning is very difficult, given that
PM$_{v}$ and PA$_{\circ}$ are determined by two completely independent phenomena. Hence, the most
likely explanation is that the velocity vector should be either parallel or perpendicular to the
rotation axis and/or PA$_{\circ}$ should be parallel or perpendicular to the fiducial plane. The
latter is a natural consequence of pulsar radio emission being excited by the soliton coherent
curvature radiation, as argued by
\citet{mgp00}, \citet{glm04}, and \citet{mgm09}, hereafter Papers I, II, and III, respectively.
As for the former possibility, it is yet to be established by the supernova explosion theory
\citep[e.g.][]{th75,c98,sp98}. However, in three cases for which X-ray information is available
\citep[Vela, B0656+14 and J0538+2817; see][]{r07}, the rotation axis is parallel to the proper
motion direction. If this can be generalized in the future, then the pulsars with $\Psi \sim
90^{\circ}$ in the bimodal distribution of $\Psi$ (see Figure 3 in Rankin 2007) represent the
X-mode observed as the primary polarization mode (PPM). Similarly those with $\Psi
\sim 0^{\circ}$ represent the O-mode observed as a secondary polarization mode (SPM).

Recently in Paper III the authors showcased highly polarized subpulses from single pulses for a
number of pulsars, in which the position angle of the linear polarization closely followed the mean
position angle traverse. They further conjectured that the observed polarization state of some
subpulses represents the extraordinary mode excited by the soliton coherent curvature radiation.
This conclusion was conditional on non fulfilment of the so called Adiabatic Walking Condition (AWC
henceforth) first introduced by \citet[CR79 hereafter]{cr79}. The conclusions of Paper III can
remain unnoticed, since it lacked strong and convincing arguments that the polarization direction
of the generated waves cannot be changed by the adiabatic walking. In this paper we will give
further arguments, both from observational and theoretical points of view, that the AWC is indeed
not satisfied in pulsar magnetospheres, under conditions required for the soliton coherent
curvature radiation emitted at frequencies much lower than the local characteristic plasma
frequency. Thus the X-mode excited by the coherent curvature radiation can freely leave the pulsar
and reach the observer. This mode represents the observationally dominant PPM. We will also present
observational evidence for the O-mode, excited by the curvature radiation. This mode dominates at
the emission region (about 6 times stronger than the X-mode)\footnote{This effect is well known for
the vacuum case \citep[e.g.][]{j75} and for the plasma environment it was generalized by
\citet{glm04}; see the last paragraph in Section 5 of their paper.}, but it cannot freely propagate through magnetospheric plasma. We will discuss a possible scenario
under which part of the O-mode can escape from the magnetosphere. This mode would then correspond
to the weaker secondary polarization mode (SPM).

\section{Importance of the single-pulse polarization properties\label{secimport}}
\label{spol}

In Figure~\ref{fig1} we show an example of the observed composite and single pulse polarization in
pulsar PSR B2045--16 \footnote{ PSR B2045--16 is a slowly rotating pulsar with period of 1.9 sec,
and its position angle curve suffers almost no distortions due to the effects of aberration and
retardation \citep[see][for a detailed discussion]{ml04}, and hence follow the RVM rather
accurately.} observed with the Giant Meterwave Radio Telescope (GMRT, see Paper III for observation
details) near Pune, India. Two highly polarized subpulses, appearing in two different single pulses
in the longitude range corresponding to the leading profile component are shown in the left panels
of the figure. Their PPA traverses correspond to the two orthogonal polarization modes. These
subpulses represent essentially single polarization modes close to 100\% polarization each. They
obviously do not suffer from any significant depolarization, and the PPA associated with the
leading profile component follow the mean position angle traverse.

Without any further modeling we are unsure about the orientation of the polarization vectors of
these modes with respect to the pulsar magnetic field line planes. We can however adopt the
strategy taken by
\citet{mrg07}, where they determine the direction of the modal PPAs with
respect to the projected magnetic field line planes by comparing the fiducial PPA and proper-motion
directions as explained in the introduction. The proper motion of this pulsar is
PM$_v$=92$^{\circ}\pm2$ and the absolute value of PPA at the fiducial longitude
PA$_{\circ}$=$-3^{\circ}\pm$5, which gives the quantity $\Psi$= (PM$_{v}$- PA$_{\circ}$) =
95$^{\circ}\pm 5$ \citep[see][]{r07,mgsjst79,mgsbt81}. Thus, based on the assumption that the
pulsar proper motion is directed along the rotation axis\footnote{Recently
\citet{nkcj12} performed a statistical analysis for 54 pulsars to test the pulsar rotation axis and
proper motion alignment, and found strong evidence for this effect, excluding that these vectors
are completely uncorrelated with $>$ 99\% confidence.}, the emerging electric field at the fiducial
point of the dominant PPA mode is perpendicular to the dipolar magnetic field line plane. As seen
in the right panel of Fig~\ref{fig1}, the fiducial point is determined by the stronger PPM and
hence this point should be associated with the perpendicular or extraordinary (X) mode. The weaker
SPM should then be associated with the parallel or ordinary (O) mode. As it was pointed out in
Paper II, this argument can be extended to the rest of the PPA traverse which follow the RVM model,
and hence the polarization state for this pulsar at each pulse phase along the PPM represents the
X-mode polarized perpendicularly to the magnetic field line planes.

As a summary of this section, let us consider a fictitious radiation mechanism that generates two
orthogonal polarization modes without being parallel or perpendicular to the magnetic field line
planes. To explain the observed polarization in Figure~\ref{fig1}, one has to invoke a propagation
effect that rotates the modes in such a way that they emerge almost purely polarized parallely or
perpendicularly to the dipolar magnetic field line planes. One of the such mechanisms was proposed
by CR79, who introduced the so-called adiabatic walking scenario. They argued, that even
chaotically oriented polarization directions will slowly rotate while the waves propagate away from
the generation region. As a result within the polarization limiting region \citep[beyond which the
waves detach from the plasma; see e.g.][]{ab86} all the chaotic polarizations will get arranged and
thus the emerging polarization will consist of the two orthogonal modes.\footnote{The polarization
of one mode is parallel to the local ${\bf k}$ and ${\bf B}$ plane and the polarization of other
mode is perpendicular to it. This local plane in general does not coincide with the plane of
magnetic field lines.} However, this does not mean that they will be polarized parallel or
perpendicular to the planes of dipolar magnetic field lines, as demonstrated in Figure~\ref{fig1}
(a collection of such pulses is presented in Figure~\ref{fig1a}). This would require unrealistic
fine tuning in every pulse longitude\footnote{It is well known that the variations of the
polarization position angle across the pulse window trace the range of dipolar magnetic field line
planes as predicted by the RVM model.}. Therefore, based on our observations, we conclude that this
radiation is not affected by the adiabatic walking and we receive the generated polarization
pattern. The only radiation mechanism that can distinguish the magnetic field line planes is the
curvature radiation excited in pulsar plasma (Paper II and III). In the next section we will
demonstrate that the AWC is not satisfied in the pulsar plasma with characteristics allowing
generation of the coherent soliton curvature radiation. Therefore the adiabatic walking does not
affect the polarization of the plasma X- and O-waves excited by this radiation mechanism.

Before closing this section, it is important to comment on the majority of subpulses that has
intermediate, low or undetectable polarization, where the PPA's do not necessarily follow the mean
PPA traverse. This property is in general expected in the currently discussed coherent curvature
radiation model (Papers I, II and III). The subpulse in this model results from incoherent addition
of a large number of independently emitting solitons. Each of these solitons excite the X- and
O-modes in the plasma, however the  modes get separated (although, in this paper we will show that
the modes do not change their state of polarization) and further the amplitude of the O-mode can be
significantly damped. Before the modes emerge from the plasma, the incoherent addition of a large
number of such waves (with random contribution of X- and O-modes both in amplitude and phase) leads
to a range of observed depolarization values (detailed considerations of that scenario has been
postponed to a separate paper). Only in certain occasions the plasma conditions are such that
either the X- or O-mode emerges as the dominant radiation mode, which is observed as close to 100\%
polarized subpulse emission. In fact the highly polarized subpulses presented in this paper and
Paper III are indeed rare (typically $<1\%$) occurrences. However, they are very important to be
studied as they can reveal the true nature of pure X- and O-modes.

\section{Plasma waves in the near magnetosphere and Pulsar emission Mechanism
\label{secpwave}}
The coherent curvature radiation has been considered as a natural emission mechanism for the
observed pulsar radiation \citep[e.g. Ruderman
\& Sutherland 1975, RS75 hereafter;][]{bb83,mp84}, although several key issues in this theory
remained unresolved for a long time. For example, the theory of formation of charged bunches
leading to curvature radiation in a plasma was not developed. Also, it was not clear how the
radiation emerges from the magnetospheric plasma. The recent attempts to build and develop a self
consistent theory of curvature radiation for pulsar radio emission has been published in Papers I,
II, and III. Under the framework of this theory, here we recapitulate the physical processes
generating the magnetospheric plasma that leads to the observed pulsar coherent radio emission.

The basic requirement of this theory is the presence of non-stationary inner acceleration region
above the polar cap.  The prototype of this region was the inner vacuum gap suggested by RS75. The
formation of their gap was based on the overestimated value, by at least an order of magnitude, of
the binding energy of iron ions in the neutron star crust \citep[][and referencies therein]{gmg03}.
As a result the accelerating potential drop along the magnetic field in the gap was very high,
exceeding 10$^{13}$ volts. In such high potential drop the backflowing particles heat the polar cap
to enormous temperatures close to $10^{7}$ K, which has been never observed in the form of intense
thermal X-rays form polar caps. Also the so called subpulse drift would be too fast as compared
with the observed drift rates \citep[][and referencies therein]{gmg03}. Therefore the original RS75
model called for a revision from the very beginning.

In the mean time, new calculations of binding energy of iron ions was published by \citet{j86} and
similar study was performed using different and more precise method by \citet{ml06,ml07}. These
calculations demonstrated that the vacuum gap could be formed only if the actual surface magnetic
field at the polar cap was close to 10$^{14}$ G, much higher than the dipolar magnetic field at the
surface of most pulsars. To preserve the idea of the inner accelerator region one has to invoke a
strong non-dipolar crust anchored surface field much stronger than the dipolar component at the
surface. It appears that such surface magnetic fields are likely to be produced in the neutron star
crust by Hall-drift instabilities \citep[and references therein]{ggm13}.

Based on these ideas \citet{gmg03} proposed the Partially Screened Gap (PSG hereafter) model for
the inner acceleration region. In this model the backflowing particles heating the polar cap cause
a thermal outflow of iron ions, which partially screen the gap potential drop. The more they
screen, the less the lower potential drop causes heating and therefore less screening at the same
time. So the thermostatic regulation will occur establishing the surface temperature at a value
close to but slightly below the so-called critical temperature (at which iron ions would be
extracted at the co-rotation limited Goldreich-Julian charge density). In the consequence the
maximum value of the potential drop will be about one order of magnitude lower as compared with the
vacuum case of RS75. This potential drop is still high enough to drive the non-stationary sparking
discharge in the PSG and produce the electron positron primary plasma similarly to that envisioned
by RS75. The presence of an additional iron ion component does not influence significantly the
production of the secondary plasma. The so-called multiplicity factor $\kappa = n_{\pm}^{s}/n_{GJ}$
(where $n_{\pm}^{s}$ is charge density of the secondary plasma and $n_{\rm GJ}$ is the
Goldreich-Julian charge density) lies in the range 100 $\leq \kappa \leq 10^5$ (for details see
\citet{am98}, also \citet{sz13} for application to the PSG model). The structure of the secondary
plasma is formed by the sparking discharge of the PSG.

\subsection{Plasma Dispersion relation at the Radio Emission Region}
Let us now consider the magnetospheric plasma waves in the radio emission region, i.e. the tube of
the open field lines consisting of electron and positron of secondary plasma at altitudes $r$ of
about 50 to 100 stellar radii, where the coherent curvature radio emission is expected to
originate. We start with the dispersion equation in the pulsar frame of reference assuming that
${\bf k}$ is along the $z$-axis and $\vartheta $ is the angle between the local magnetic field
${\bf B}$ and the wave vector ${\bf k}$ \citep[see ][KMM91 hereafter]{kmm91}. The distribution
function of the secondary plasma is close to a gaussian distribution, centered at $\gamma =$
$\gamma_{s}\gg1$, having a spread of $\Delta \gamma /\gamma_{s}\ll 1$ (Paper II, Szary 2013). At
the altitudes $r=(50 - 100)\times10^6$ cm the fractional altitude $\Re\equiv (r/R_{\rm LC})=2\pi
r/(cP)$ is for typical pulsars small ($\Re= (0.01 - 0.1)P^{-1}$) and hence the ratio
\be
\frac{\omega_{p}}{\omega_{B}}=2.8\times10^{-4}\kappa^{0.5}P^{0.75}{\dot
  P}_{-15}^{-0.25}\Re^{1.5}\ll 1, \label{wpwb}
\ee
where $\omega_p^2=4\pi e^2 \kappa n_{GJ}/m_e$ is the plasma frequency, $\omega_B=eB/m_{e}c$ is the
cyclotron frequency, $m_e$, $e$ and $c$ correspond to the mass of the electron, charge of the
electron and speed of light, respectively (in cgs units).

Using the conditions stated above we can neglect all terms containing $\omega_{B}^{-1}$ and hence
the dispersion relation can be written in the form
\be
\left( k^{2}\frac{c^{2}}{\omega^{2}}-\varepsilon_{11}\right) E_{1}-\varepsilon_{13}E_{3}=0,
\label{eps1}
\ee
\be
\left(k^{2}\frac{c^{2}}{\omega^{2}}-1\right) E_{2} =0,
\label{eps2}
\ee
\be
\varepsilon _{31}E_{1}+\varepsilon _{33} E_{3} =0.
\label{eps3}
\ee
Here $E_{1}$, $E_{2}$ and $E_{3}$ are the components of electric
fields along $x$,$y$ and $z$ axes ($y-$axes is perpendicular to the
plane of ${\bf k}$ and ${\bf B}$), $ \varepsilon_{11}
=1-I\sin^{2}\vartheta$, $\varepsilon_{13} =\varepsilon_{31} =
I\sin\vartheta\cos\vartheta$, $\varepsilon_{12}
=-\varepsilon_{21}=\varepsilon_{23}=-\varepsilon_{32} =0$,
$\varepsilon_{33} =1-I\cos^{2}\vartheta$, where $I$ can be expressed
as:
\be I\equiv \frac{1}{2}\sum \omega_{p\alpha }^{2}\int
\frac{dp}{\gamma^{3}}\frac{f_{\alpha }}{\left( \omega -kv\cos\vartheta
  \right)^{2}},
\label{I}
\ee
$\alpha$ denotes sum over electrons and positrons, $\omega_{p}^{2}=4\pi e^{2}n_{\pm}^{s}/m_{e}$,
$p$ and $v$ are the dimensionless momentum and velocity of the particles along the magnetic field
(see KMM91). Solving the above equations, we get the dispersion relation
\be
1-\frac{\omega^{2}}{k^{2}c^{2}}=0
\label{drxmode}
\ee
for the X-mode, which is linearly polarized along $y$-axes, while for the O-mode (polarized in
$xz$-plane) the dispersion curve is a solution of the following equation
\be
\frac{1}{2}\left(1-\frac{k^{2}c^{2}}{\omega ^{2}}\cos ^{2}\vartheta \right)
\sum \omega_{p\alpha}^{2}\int \frac{dp}{\gamma ^{3}}
\frac{f_{\alpha }}
{\left(\omega -kv\cos \vartheta \right) ^{2}}=1-\frac{k^{2}c^{2}}{\omega ^{2}}.
\label{epsomode}
\ee
Let us note, that under the condition (\ref{wpwb}) we can assume that the distribution functions
for both electrons and positrons are identical. The difference between these functions is important
for the soliton coherent curvature radiation as the slowly varying charge-density is proportional
to $\Sigma e_{\alpha}\omega_{p\alpha }^{2}$ and vanishes if the distribution functions are
identical (see Equation (A19) in Paper I). On the other hand, those components of the linear
permittivity tensor, i.e. $\varepsilon_{12}$, $\varepsilon_{21}$, $\varepsilon_{23}$ and
$\varepsilon_{32}$, which have non-zero value in the case of non-identical distribution functions
(see, e.g. KMM91), are proportional to $\omega_{B}^{-1}$ and are negligibly small (see
Equation~(\ref{wpwb})). Thus, taking into account $\Delta \gamma /\gamma_{s}\ll 1$ the distribution
functions can be described by the delta-function at $\gamma_{s}$ i.e. $f_{\pm }\sim \delta \left(
p-p_{s}\right) $. For the purpose of our discussion this is a very reasonable assumption, as we are
interested in the average features of the plasma, as for the peculiarities of plasma distribution
they are irrelevant. Then, using $v\approx 1-0.5\gamma ^{-2}$ Equation (\ref{epsomode}) can be
reduced to the following form:
\be
1-\frac{k^{2}c^{2}}{\omega ^{2}} =\frac{\omega_{p}^{2}}{\gamma_{s}^{3}}
\frac{\left(
  1-\frac{k^{2}c^{2}}{\omega^{2}}\cos ^{2}\vartheta \right)}{\left(
  \omega -k\left( 1-\frac{1}{2\gamma_{s}^{2}}\right) \cos \vartheta
  \right) ^{2}}, \label{eq_omode}
\ee
where $\gamma_{s}=\left(1+p_{s}^{2}\right) ^{0.5}$. Equation~(\ref{eq_omode}) describes the
ordinary modes (i.e. modes polarized in the plane of ${\bf k}$ and {\bf B}) of the magnetized
electron-positron plasma and has two possible solutions, which correspond to two branches: O-mode
and L-mode. In the conditions under consideration (see Section 3.2 for details), i.e. $\vartheta
\ll 1$ and $\omega$ being in the range from 0 up to values that does not exceed $\omega_0$
significantly (see Equation~(\ref{omega0}) below), the waves have following features: the L-mode is
almost longitudinal, while the O-mode is almost transverse. Equation~\ref{eq_omode} cannot be
solved analytically, and to demonstrate its basic properties the schematic representation of
solutions are shown in Figure~\ref{fig2}. In the frequency region $\omega \gg \omega_{\circ}$ the
polarization characteristics of the L- and O-modes change, the L-mode becomes almost
electromagnetic while the O-mode turns to be almost longitudinal and as a subluminal wave, it
undergoes strong Landau damping (see Section 3.5).

If $\vartheta =0$ (propagation along the local magnetic field direction) Equation~(\ref{epsomode})
has two independent solutions. The purely transverse solution coincides with X-mode, while L-mode
coincides with the purely longitudinal Langmuir wave, the dispersion equation of which is obtained
by dividing Equation~(\ref{epsomode}) by Equation~(\ref{drxmode}). Two values of wave frequency can
be distinguished. The first point is defined by $k=0$ for which the corresponding frequency is
$\omega =\omega _{1}=\gamma_{s}^{-1.5}\omega_{p}$. The second point is defined by the condition
$\omega=kc$, i.e. the phase velocity equals $c$ for which the corresponding frequency is
\be \omega =\omega_{\circ}=2{\sqrt\gamma_{s}}\omega_{p}.
\label{omega0}
\ee
In this paper we use the term characteristic frequency for $\omega_{\circ}$. It is important to
note that Langmuir waves cannot be exited if $\omega_1<\omega<\omega_0$, since the phase velocity
exceeds the speed of light (superluminal waves). On the other hand, in frequency region
$\omega>\omega_0$ the phase velocity is less than the speed of light (subluminal waves), thus the
two-stream instability can develop, provided that the plasma distribution function has a proper
shape \citep[see][for details]{am98,lmmp86}.
\subsection{Domain of plasma parameters}
\label{secppar}
Let us overview basic parameters of the secondary plasma flowing along the tube of open field lines
in the observer's frame of reference. We will represent them in terms of the basic pulsar
parameters $P$ (in sec) and ${\dot P}={\dot P}_{-15}\times 10^{-15}$ (s/s), and as a function of
the fractional altitude $\Re=r/R_{\rm LC}$ (see previous section). The cyclotron frequency $\nu_B$
in the local magnetic field $B$ can be expressed in GHz as,
\begin{equation}
\nu_B = \frac{\omega_B}{2\pi\gamma_s} = 5.2 \times 10^{-2} \frac{1}{\gamma_s} \left(\frac{{\dot
P_{-15}}}{P^5}\right)^{0.5} \Re^{-3}. \label{nu_B}
\end{equation}
The characteristic  plasma frequency $\nu_{\circ}$ in GHz can be written as:
\begin{equation}
\nu_{\circ} = \frac{\omega_0}{2\pi}= 2\times 10^{-5}\kappa^{0.5} \sqrt{\gamma_s}\left(\frac{{\dot
P_{-15}}}{P^{7}}\right)^{0.25}\Re^{-1.5}, \label{nu_0} \ee with $\omega_0$ given by Equation
(\ref{omega0}). As the third parameter we present the characteristic frecuency of the soliton
curvature coherent radiation, wich corresponds to the maximum of the power spectrum $\nu_{\rm cr}$
(see Figure~2 in Paper II),
\begin{equation}
\nu_{\rm cr} = 1.2 \frac{c {\Gamma}^3}{2 \pi \rho}= 0.8 \times 10^{-9}
\frac{{\Gamma}^3}{P}\Re^{-0.5}, \label{nu_cr}
\end{equation}
where $\rho$ is the radius of curvature of the dipolar magnetic field lines and $\Gamma$ is the
Lorenz factor of the emitting soliton, for which the oppening angle of the radiation cone
$\vartheta\sim\Gamma^{-1}$.

Figure~\ref{fig3} summarizes these basic plasma parameters for a canonical pulsar with $P$= 1 sec
and ${\dot P}_{-15}$=1, in the observer's frame of reference as a function of fraction of the light
cylinder distance, put into the context of the observed pulsar radio emission. The abscissa axis
originates at the altitude of about 5 stellar radii, above which the crucial two stream instability
can develop \citep{am98}\footnote{If in Equation (63) by \citet{am98} we assume reasonably low
values for the average Lorentz factor of plasma particles $\gamma_p/10^2\sim (0.5 - 0.7)$ and for
the half-width of a plasma cloud $\Delta\Psi /(5\times 10^4)\sim (0.2 - 0.4)$ we find that a
corresponding distance from the center of the star $r$ ranges from a few to several stellar radii.
Thus the altitude of about 5 stellar radii seems to be quite a good approximation for the minimal
altitude at which the two-stream instability can develop.}. The uppermost red line represents the
cyclotron frequency $\nu_{B}$ (given by Equation~(\ref{nu_B})), where $\gamma_s = 200$ was chosen
as an average Lorentz factor of the secondary plasma (note that in the scale used in this plot,
changing $\gamma_s$ between 100 to 1000 will not move noticeably the red line). This line
demonstrates that any radiation with frequency higher than $\nu_{B}$ will be damped by means of
cyclotron resonance. The next pair of green lines represent the plasma frequency $\nu_{\circ}$
(given by Equation~(\ref{nu_0})) calculated for $\gamma_s = 200$ and two limiting values of $\kappa
= 100$ and $10^4$. Two grey lines parallel to the x-axis, represent the canonical range of the
observed pulsar radio emission (between 30 MHz to 30 GHz), with a dashed line corresponding to 1
GHz. The remaining two dark blue lines represent characteristic frequencies $\nu_{cr}$ of the
curvature radiation (given by Equation~(\ref{nu_cr})) calculated for the radius of curvature of
dipolar field lines ($\rho \sim 10^8$ cm) and for two values of the Lorentz factor of the charged
soliton $\Gamma = 600$ for the upper solid line and $\Gamma = 400$ for the lower dashed line (see
Paper II for details).

We can now briefly discuss how Figure~\ref{fig3}  will alter for different pulsar periods. The
frequencies $\nu_{\circ}$ and $\nu_{B}$ are much more sensitive to changing $P$ than the
characteristic frequency of the curvature emission $\nu_{\rm cr}$, which depends on period in the
same way as the fractional distance. Therefore, the soliton curvature radiation can always develop
in the proper region of altitudes. i.e about 50-100 stellar radii, irrespective of the value of
$P$.
\footnote{However for very long periods the available potential drop could not be high enough to power the primary beam,
which leads to "death" of radio pulsars. It is also worth mentioning, that for very old pulsars the crustal strong
surface field may not be supported by the Hall drift instability. Absence of the crustal field leads to switching off
the creation of dense electron-positron plasma, necessary for a mechanism of the coherent pulsar radio emission
\citep{ggm13}.}

Figure~\ref{fig3} corresponds to the total power and contains no information about the polarization
properties of the observed radiation. These properties will be considered in Section (\ref{secAWC})
in the context of possible propagation effects.
\subsection{The mechanism of coherent radio emission}
\label{secmech}
Let us now briefly review the spark-associated soliton model of the coherent curvature radiation
from pulsars (for details see Paper I and references therein). Like any plasma model for pulsar
radiation this model is also based on development of some plasma instabilities. The only plasma
instability that can arise at altitudes lower than 10\% of the light cylinder is the two-stream
instability \citep[Lominadze et al. 1986; KMM91;][]{am98}, as all other instabilities are suspended
by the strong magnetic field. The two-stream instability is a result of the effective energy
exchange between particles and waves, that can occur if the phase velocity of waves ($\omega/k$) is
near to the velocity of resonant particles ($v_r$), i.e. the resonant condition $(\omega - kv_r)=0$
is satisfied. As it has been already mentioned, the resonance is possible in frequency region
$\omega>\omega_0$. However, one should realize that amplification of waves occurs only if there is
an excess of particles with velocities larger than the phase velocity of waves near the resonant
point. In the opposite situation the waves will be damped (the Landau damping). Thus, the plasma
distribution function should have a proper shape
\citep[e.g.][]{am98} for a two-stream instability development. Such conditions can naturally be
realized if a plasma is produced via non-stationary gap discharge. The spark discharge timescale in
the PSG is a few tens of microseconds and this results in overlapping of successive clouds of
outflowing secondary plasma. Each elementary spark-associated plasma cloud has a spread in momentum
and the overlapping of particles with different momentum leads to two stream instability in the
secondary plasma cloud \citep[RS75;][]{usov87, am98}. This triggers strong Langmuir turbulence in
the plasma and if this turbulence is strong enough, the waves become modulationally unstable. The
unstable wave packet described by the nonlinear Schr\"{o}dinger equation leads to formation of a
quasi-stable nonlinear solitary wave, i.e. a soliton \citep{pm80}. The longitudinal (along ${\bf
k}$) size of the soliton should be much larger than the wavelength of the linear Langmuir wave.
Also it has to be charged to be able to radiate coherent curvature emission. Thus, the soliton
bunch has to be charge separated, which can be caused either by difference in the distribution
function of electrons and positrons, or by admixture of iron ions in the secondary plasma or by
both these effects. A sufficient number of charged solitons is formed which can account for the
observed radio luminosity in pulsars (Paper I). The wavelength of the emitted waves should be
longer than the longitudinal size of the soliton $\Delta$. This is the necessary condition for the
coherency of a curvature radiation process. Thus, the frequencies plotted in Figure (\ref{fig3})
should obey the following constraints
\be
\nu_{\rm cr}<\frac{c}{\Delta}\ll
\nu_{\circ} \ll
\nu_{B}. \label{condition}
\ee
It is clearly seen from Figure (\ref{fig3}) that the observed pulsar radiation cannot be generated
at altitudes exceeding 10\% of the light cylinder radius (practically the radio emission region
should be contained between one to several percent of $R_{LC}$; see dashed area in Figure
(\ref{fig3})). This conclusion is based purely on the properties of plasma and the emission
mechanism. And it corresponds perfectly to the other limits on emission heights obtained from
observations and geometrical considerations
\citep[e.g.][]{cor78,bcw91,r93,kg97,md99,gg03,ml04,kmgkga09}.

\subsection{Adiabatic walking condition}
\label{secAWC}
While studying rotation of polarization planes of X- and O-modes propagating in the pulsar
magnetospheric plasma CR79 introduced the AWC in the form of
\be
\left\vert \frac{1}{k}\frac{\partial }{\partial x}\phi \right\vert \ll \left\vert \Delta
N\right\vert,
\label{eqawc}
\ee
where $\Delta N$ is a change of $N$ occurring during propagation of waves, $\phi$ is the linear
polarization angle (dimensionless) of the mode or some similar dimensionless parameter, which
defines the mode polarization and $1/k \equiv 2\pi \lambda$, where $\lambda$ is the
wavelength\footnote{This is the sufficient condition of CR79 (their Eq. 2). The necessary
(adiabatic) condition of CR79 (their Equation (1)) is always satisfied, if plasma properties vary
slowly enough.}. Here $N=N^{(o)}-N^{(x)}$ is the difference between refractive indices of O- and
X-mode:
$$N^{o}=\frac{kc}{\omega}|_{\rm O-mode}\ {\rm and}\ N^{x}=\frac{kc}{\omega}|_{\rm X-mode}.$$
Remembering that the polarization vector of waves is either tangent or normal to $\bf k$ and $\bf
B$ plane, for a change of the dimensionless parameter $\phi$ we can choose $\Delta
\phi=\Delta\vartheta / \vartheta$, where $\vartheta \sim 1/\Gamma$. It expresses the fact that the change
of angle $\vartheta$ between $\bf k$ and $\bf B$ during propagation causes a corresponding change
of the polarization vector. Thus, the polarization plane of a wave would rotate by the angle about
unity ($\Delta \phi \sim 1$) while it propagates the distance $\Delta l=\Delta\vartheta
\rho$, if the AWC was fulfilled (see also Paper II).

In our formalism, as seen in Equation (\ref{drxmode}), the refractive index $N^{(x)}=1$ for all
values of $k$. To obtain $N^{(o)}$ we need to solve Equation (\ref{eq_omode}), which, however,
cannot be achieved analytically. But we can make following very reasonable assumption:
\be \frac{k^{2}c^{2}}{\omega_{\circ}^{2}}\sin ^{2}\vartheta \ll 1
\label{cond_omode}
\ee
This condition is consistent with Equation (\ref{condition}), which assumes that in the radio
emission region the frequency of emitted waves ($\nu_{\rm cr}=kc/2\pi$) should be less than
$\nu_{\circ}=\omega_{\circ}/2\pi$. As $\sin\vartheta$ cannot exceed unity, the above condition is
always valid for emitted waves. The solution of Equation~(\ref{eq_omode}) for the O-mode under the
condition expressed above is
\be \omega=kc\cos \vartheta \left(
1-\frac{1}{2}\frac{k^{2}c^{2}}{\omega_{\circ}^{2}}\sin ^{2}\vartheta
\right).
\label{solution_omode}
\ee
Thus it is straightforward to show that
\be
N^{(o)}=\frac{kc}{\omega }|_{\rm O-mode}=\cos^{-1}\vartheta \left(
1+\frac{k^{2}c^{2}}{2\omega_{\circ}^{2}}\sin ^{2}\vartheta
\right). \label{ref_ind_omode} \ee
Then assuming $\vartheta \ll 1$, which is very reasonable assumption for the relativistic plasma,
we obtain that
\be
N=N^{(o)}-1=\frac{1}{2}\vartheta ^{2}\left(
1+\frac{k^{2}c^{2}}{\omega_{\circ}^{2}}\right).
\label{ref_ind_omode_1}
\ee
In order to examine validity of the AWC we can use $\left\vert \partial \phi/\partial x \right\vert
\approx \left\vert \Delta \phi /{\Delta l}\right\vert=1/\vartheta\rho$ and $\left\vert \Delta N\right\vert
=\left\vert \vartheta \Delta \vartheta \right\vert $, therefore the AWC can be rewritten in the
following form
\be \vartheta^{3}\left(
\frac{\Delta \vartheta }{\vartheta }\right) \gg \frac{1}{\rho }
\frac{\rho }{\Gamma^{3}}.
\ee
Let us note that $\Delta\vartheta/\vartheta\sim 1$ (meaning that AWC is fulfilled) only if
$\vartheta ^{3}\gg \Gamma^{-3}$. However, $\vartheta \approx \Gamma^{-1}$, thus we can safely
conclude that the AWC is not satisfied under our assumptions. Therefore, the adiabatic walking
cannot affect the polarization of waves excited by the soliton curvature radiation.

It may be interesting to figure out why we have found just an opposite conclusion to that obtained
by CR79. The main difference comes from the condition $k^{2}c^{2}=\omega^2\ll\omega_{\circ}^{2}$
which must be satisfied in the case of soliton curvature radiation (Paper I). In the case of the
classical curvature radiation model of RS75 that was used by CR79 the characteristic frequencies
should obey the following condition $k^{2}c^{2}=\omega^2\approx \omega_{\circ}^{2}$ (since their
bunches are formed by linear Langmuir waves).

In the frequency range $\omega\sim\omega_{\circ}$ (which is not allowed in the soliton mechanism,
Paper I) the dispersion curve of the O-mode deviates significantly from the X-mode dispersion line
$\omega =kc$, as compared with the frequency range of soliton curvature emission
$\omega\ll\omega_{\circ}$ (marked by the shadowed area in Figure~\ref{fig3}). It is difficult to
demonstrate these differences in Figure~{\ref{fig2}} and therefore we present the numerical
solution for $\left\vert N \right\vert$ in Figure~\ref{fig4}, obtained for three values of
$\vartheta$. These figures are equivalent to each other since Figure~\ref{fig4} represents simply
the difference between the O-mode and the X-mode in Figure~\ref{fig2}. As we can see from
Figure~\ref{fig4}, below $kc/\omega_{\circ} \sim 0.1$, $\left\vert N \right\vert$ stays unchanged,
so $\Delta N \sim 0$ and the AWC cannot be satisfied. This is the range of soliton coherent
curvature radiation considered in this paper as the viable pulsar radio emission mechanism. Beyond
this range $\left\vert N
\right\vert$ becomes varying so $\Delta N$ can attain relatively large values and therefore the AWC
can be satisfied. This range was considered by CR79 who used RS75 radiation model, in which charged
bunches formed by linear Langmuir waves emitted coherent curvature radiation. However this
radiation mechanism should not be realized physically as argued for the first time by
\citep[Lominadze et al.(1986)]{lmmp86} and summarized in Paper I. It is worthwhile to
recapitulate these arguments here. The emission of waves (by means of curvature radiation) with
frequency close to the local plasma frequency $\omega_{\rm cr}=\omega_{\circ}$ is impossible,
because one cannot fulfill the following two conditions simultaneously: (1) the timescale of the
radiative process must be significantly shorter than the plasma oscillation periods, which means
$\omega_{\rm cr}\gg\omega_{\circ}$, and (2) the linear characteristic dimension of the bunches must
be shorter than the wavelength of the radiated wave, which means $k_{\rm cr}\ll
k_{\circ}=\omega_{\circ}/c$. On the other hand $\omega_{\circ}=k_{\circ}c$ and $\omega_{\rm
cr}=k_{\rm cr}c$. Thus it is impossible to satisfy the above two conditions simultaneously and
therefore bunching associated with high-frequency Langmuir plasma wave cannot be responsible for
the coherent pulsar radio-emission (see also \citet{mg99}).

In the soliton model of coherent curvature radiation considered in this paper, the soliton-like
bunches are formed due to the non-linear evolution of Langmuir wave packets and thus the sizes of
the bunches are naturally much larger than the Langmuir wave-length. Hence the condition $k_{\rm
cr}^{(\rm sol)}\ll k_{\circ}$ is always fulfilled (here the superscript ``sol'' corresponds to
soliton.) At the same time the soliton lifetime must be much longer than the period of Langmuir
waves ($\Delta t\gg2\pi /\omega_{\circ}$). However, $\omega_{\rm cr}^{\rm(sol)}\lesssim \Delta
t/2\pi$ or $\omega_{\rm cr}^{\rm(sol)}\ll \omega_{\circ}$. As we have shown, this condition implies
that the AWC cannot be satisfied for the plasma modes excited by the soliton curvature radiation
mechanism. Thus, the excited plasma waves can retain their initial polarization, while they
propagate through the magnetospheric plasma.

\subsection{O-mode properties}
\label{secomode}

The group velocity of plasma waves describes the velocity and the direction of energy transfer and
is defined as
\be
{\bf v}_{g}={\bf i}\frac{\partial \omega}{\partial k_{\parallel }}+{\bf j} \frac{\partial \omega
}{\partial k_{\perp }}.
\ee
Here ${\bf i}$ and ${\bf j}$ are unit vectors directed along and across the external magnetic field
vector. As the dispersion law of X-mode is $\omega =kc$, the group and phase velocities of X-mode
are equal to each other, thus $v_{g}=c$ and ${\bf v}_{g}\parallel {\bf k}$.

The dispersion law of O-mode can be expressed as $\omega =
k_{\parallel}c(1-k_{\perp}^2c^2/2\omega_0^2)$, thus the group velocity of O-mode can be calculated
as
\be
{\bf v}_{g}={\bf i}c\left( 1-\frac{k_{\perp}^2c^2}{2\omega_{\circ}^2}\right) -{\bf
j}\frac{k_{\parallel }k_{\perp }c^{3}}{2\omega_{\circ}^2}.
\ee
Taking into account that both $k_{\parallel}$ and $k_{\perp }$ are much less than
$\omega_{\circ}/c$, it is straightforward to obtain ${\bf v}_{g}\approx {\bf i}c$. Thus the group
velocity of O-mode is directed along the external magnetic field i.e. the O-mode is ducted along
${\bf B}$ preserving direction of the wave vector and eventually decays as a result of the Landau
damping \citep{ab86}. Thus under normal conditions the O-mode cannot escape from the magnetosphere.

Let us consider under what  special conditions a fraction of the generated O-mode could escape and
reach the observer. Such a possibility can be illustrated by means of schematic Figure \ref{fig2}
presenting wave modes in strongly magnetized plasma. As it has been already stated in Section 2,
solution of Equation (\ref{eq_omode}) corresponds to two types of waves: superluminal (L-mode) and
subluminal (O-mode). In the case of oblique (nonzero $\vartheta$) propagation both of them are
mixed transverse-longitudinal waves \citep[see, e.g.][KMM91]{ab86}. Which of those two feature
actually dominates depends on the value of ratio $E_{1}/E_{3}$ (recall that $E_{3}$ is along
$\bf{k}$ and $E_{1}$ is in the plane of $\bf{k}$ and $\bf{B}$). If $E_{1}/E_{3}\gg1$ then the
O-mode (polarized in the $yz$ plane) is mostly transverse and thus almost electromagnetic in
nature. In the opposite case if $E_{1}/E_{3}\ll1$ the waves are longitudinal plasma oscillations,
thus they are non-electromagnetic in nature. Therefore, if one can demonstrate that
$E_{1}/E_{3}\gg1$ in the radio-emission region then the L-mode (provided it can be excited by some
physical mechanism\footnote{It worth mentioning that the L-mode cannot be excited directly by
curvature radiation as it is suppressed by Razin's effect (see Paper II).}; some possibility will
be discussed later) can escape as the observed orthogonal mode with the electric field lying in the
plane of curved magnetic field lines.

Let us now estimate the value of $E_{1}/E_{3}$ within our scenario.
It follows from Equation (\ref{eps1})
\be
\frac{E_{1}}{E_{3}}=\frac{\epsilon_{31}}{k^2c^2/\omega^2-\epsilon_{11}}=
\frac{I\sin\vartheta\cos\vartheta}{I\sin^2\vartheta-\left(1-k^2c^2/\omega^2\right)}.
\label{ratio_1}
\ee
Substituting $I$ from Equation (\ref{eq_omode}), Equation (\ref{ratio_1}) becomes
\be
\frac{E_{1}}{E_{3}}=-\left( \frac{\omega^{2}}{k^{2}c^{2}}\right) \frac{\sin\vartheta
}{\left(\frac{\omega^{2}}{k^{2}c^{2}}-1\right)\cos\vartheta }.
\label{ratio_2}
\ee
For the superluminal L-mode $\omega>kc$ and under small $\vartheta$ approximation
(remembering that $\vartheta\gamma_{s}>1$) the solution of Equation (\ref{eq_omode}) is
\be
\frac{\omega }{kc}=1+\frac{1}{2}\frac{1}{\vartheta^{2}\gamma_{s}^{4}}
\left(\frac{\omega_{\circ}^{2}}{k^{2}c^{2}}\right),
\label{Lmodesol}
\ee
which is valid provided that
\be
\frac{1}{\vartheta^{2}\gamma_{s}^{4}}
\left(\frac{\omega_{\circ}^{2}}{k^{2}c^{2}}\right)\ll1.
\ee
Incorporating Equation (\ref{Lmodesol}) into Equation
(\ref{ratio_2}) we obtain
\be
\frac{E_{1}}{E_{3}}=\frac{k^{2}c^{2}}{ \omega_{\circ}^{2}}2\vartheta
^{3}\gamma_{s}^{4}\gg1.
\label{ratio_3}
\ee
As it has been mentioned earlier, this condition means that the L-mode is almost electromagnetic in
nature.

Below we describe a possible scenario how this electromagnetic L-mode can be excited. The shaded
region in Figure~\ref{fig2} represents the frequency range of generated waves by means of soliton
coherent curvature radiation. Let us recall that our pulsar model is conditional on a
non-stationary sparking discharge of the inner accelerating region (PSG). Therefore the resultant
secondary plasma must be inhomogeneous in space. In such plasma there should be many regions with
steep density gradients along as well as across the magnetic field lines. If the O-mode reaches the
steep gradient of plasma density, where the number density (and hence the characteristic frequency
$\omega_{\circ}$) changes rapidly, then the L-mode curve moves towards the shaded region (shown
schematically in Figure \ref{fig2}). Under such condition the O-mode can be linearly coupled with
the L-mode and escape as electromagnetic waves (e.g. Arons \& Barnard 1986; KMM91). The L-mode also
preserves the polarization properties of the ordinary modes since both the L- and O-modes should be
linearly polarized in the same plane, which is orthogonal to the polarization plane of X-mode. Thus
the observed secondary polarization mode (i.e. the ordinary mode) will be polarized in the plane of
curved magnetic field lines. It is worth realizing that most of the generated power is contained in
the O-mode (about 6 times stronger than the X-mode; see footnote 2). Therefore, only small fraction
of this mode has to escape to explain the observed level of the weaker secondary polarization mode.

Finally, it is important to emphasis that the escape of L-mode should happen before the O-mode is
damped. This problem however is beyond the scope of discussion of this work, and will be addressed
in a forthcoming paper.

\section{Conclusions}
\label{con}

In this paper we have shown the observational evidences that the linearly polarized waves emerge
from the magnetosphere either parallel or perpendicular to the magnetic field line planes. We
associate these waves with the X- and O-modes excited by soliton coherent curvature radiation in
the secondary plasma. The above conclusion is true if the waves emerge from the magnetosphere as
they are generated, preserving their polarization properties. In other words the AWC should not
hold in the pulsar magnetosphere within the soliton coherent curvature radiation model.

We have demonstrated that properties of the pulsar plasma as well as features of the soliton
coherent curvature radiation emission mechanism provide the proper conditions for the generation of
pulsar radio emission at the altitudes well below 10\% of the light cylinder radius (see
Figure~\ref{fig3} and section~\ref{secppar} and \ref{secmech}), which is in a good agreement with
observations. Then in section~\ref{secAWC} we proceeded to show that in this emission region the
difference between the refractive indexes of O- and X-modes $N=N^o-N^x$ is negligibly small (see
Figure~\ref{fig4}) for the frequency range $\omega\ll\omega_{\circ}$, for which the soliton
coherent curvature radiation can operate. This implies that $N \sim 0$ and hence the AWC given by
Equation~(\ref{eqawc}) cannot be satisfied. It is important to note that our conclusion differs
from that of CR79 since they calculated the AWC condition at $\omega \sim \omega_{\circ}$, while
for our case $\omega \ll
\omega_{\circ}$. The excited waves can hence retain their initial polarization as they propagate in
the magnetospheric plasma. The X-mode, which is electromagnetic in nature, escapes from the pulsar
magnetosphere and represents the PPM highly polarized subpulses showcased by Paper III. Although
observationally there exist highly polarized pulses in the SPM, they cannot be related directly to
the O-mode, as under normal circumstances this mode gets damped and cannot escape. As a possible
interpretation of the SPM we suggest in Section~\ref{secomode} that if there are steep density
gradients in the radiation excitation region, then the O-mode couples to the L-mode and can emerge
as the secondary polarization mode polarized in the planes of the curved magnetic field lines.

\begin{acknowledgements}

We thank the anonymous referee for his critical comments which helped us to improve the manuscript
significantly. This paper was financed by the Grant DEC-2012/05/B/ST9/03924 of the Polish National
Science Center. JG and GM are grateful to the Inter University Centre for Astronomy and
Astrophysics, Pune India, for their kind hospitality during their stay there where the important
part of this work was completed. We thank Dipankar Bhattacharya for giving extensive support during
this work. We thank Rahul Basu for critical reading of the manuscript and comments. This paper used
data from the Giant Meterwave Radio Telescope, and we thank the staff of the GMRT for their
technical support provided during these observations. GMRT is run by the National Centre for Radio
Astrophysics of the Tata Institute of Fundamental Research
\end{acknowledgements}

\clearpage

\begin{figure*}
\begin{center}
  \includegraphics[height=75mm,width=120mm]{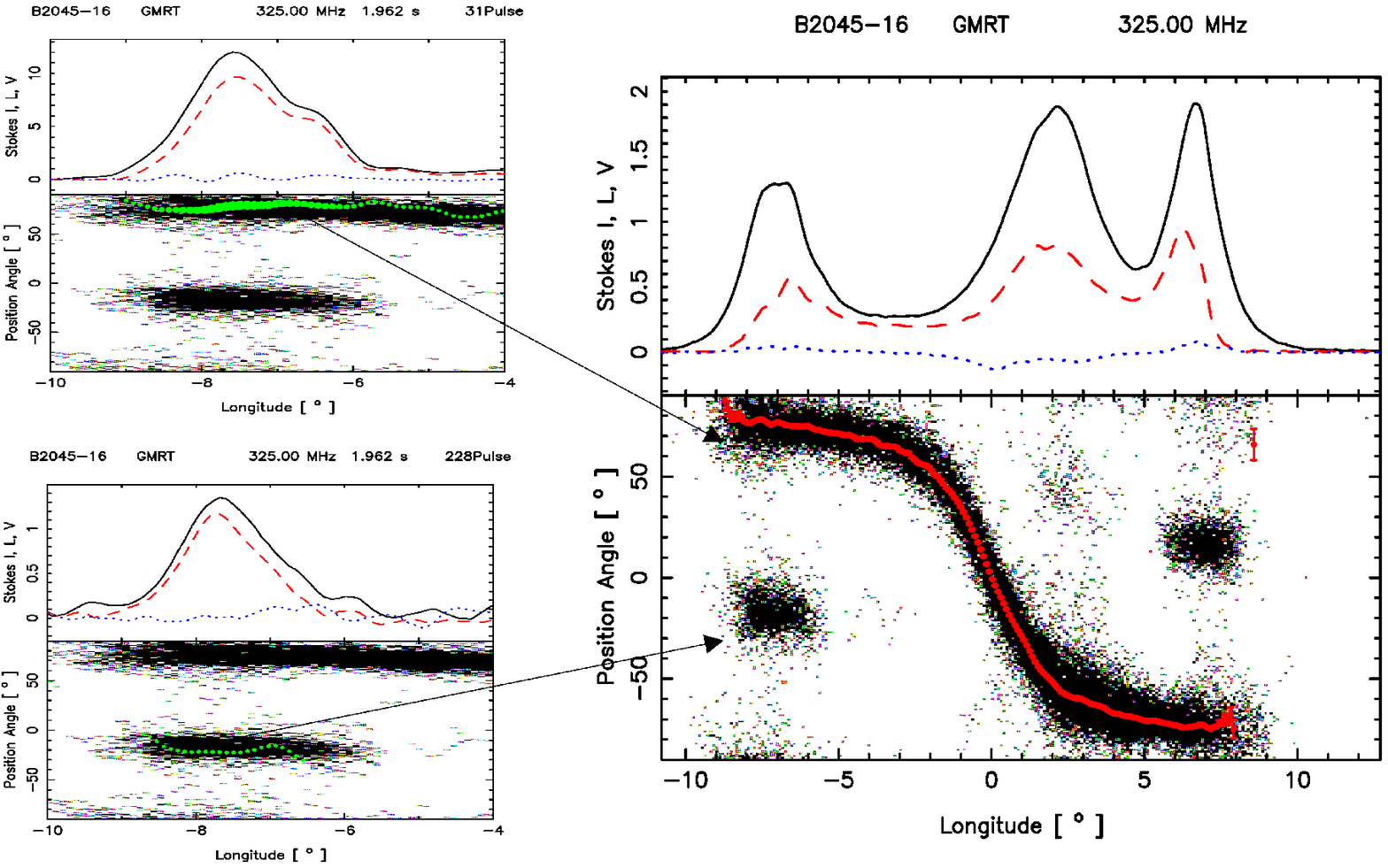}
    \caption{The  right hand figure in this plot shows the average pulse profile
             of PSR B2045--16. The top panel of this plot shows the total intensity in
             black, linear polarization in dashed red lines and circular polarization in
             dotted blue lines. In the bottom panel the red curve is the average PPA
             track computed using the average Stokes $U$ and $Q$, and the dotted points are
             the PPA histograms which are obtained by overplotting the
             PPA of every single pulse.
             The average PPA follow the more frequently occurring PPA track
             known as the primary polarization mode (PPM) and the other one is called the secondary
             polarization mode (SPM). The two plots in the left hand
             panel show two highly polarized subpulses occurring close to the leading profile component.
             The top subpulse is dominated by PPM and the bottom one is dominated by the SPM. In these subpulses the
             corresponding PPA are displayed in green, overlayed on the same dotted histograms as
             displayed in the right panel for easy reference. These histograms are exactly the
             same in all the panels of Figures~\ref{fig1} and \ref{fig1a}, as they represent all position
             angle data taken for this pulsar. It is important to realize that these two subpulses are by no means exceptional.
             An exemplary collection of similar cases can be viewed in Figure~\ref{fig1a} }
    \label{fig1}
\end{center}
\end{figure*}

\clearpage

\begin{figure*}
\begin{center}
  \includegraphics[height=180mm,width=70mm]{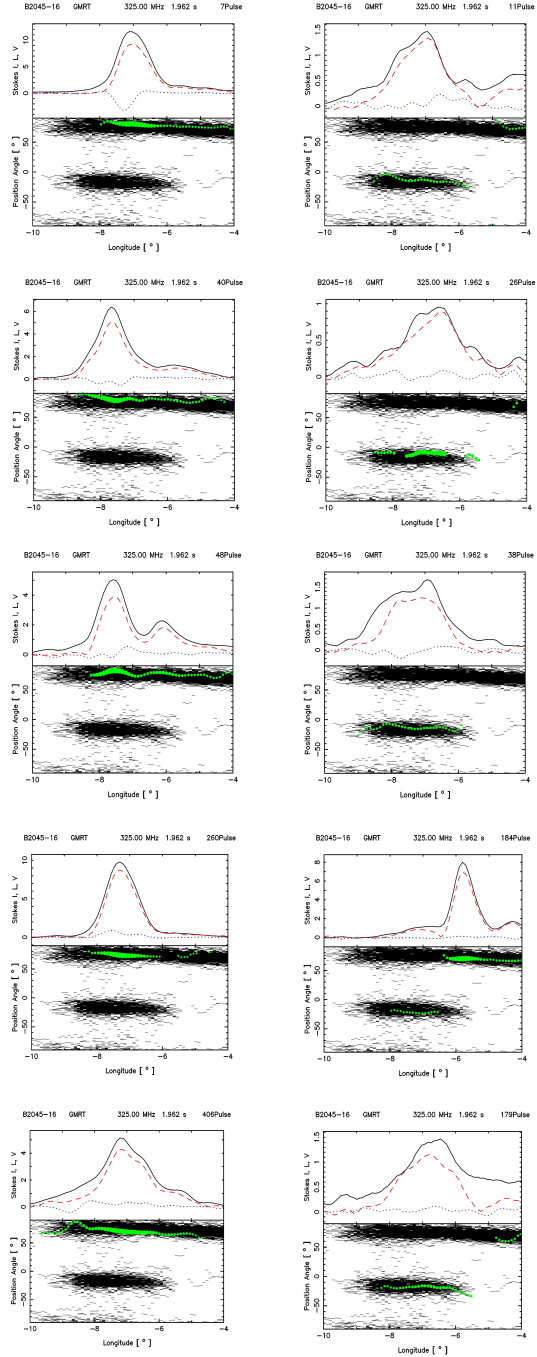}
     \caption{This Figure shows 5 more examples of highly polarized subpulses shown the same way as
             in Figure~\ref{fig1}. The left panels show stronger subpulses dominated by the primary
             polarization mode (PPM). The right panels show weaker subpulses dominated by the secondary
             polarization mode (SPM). However, in these panels there are also subpulses dominated by PPM,
             although in all presented individual pulses the single subpulses are always polarized in one of the modes.
             For further details see section~\ref{spol}.}
  \label{fig1a}
\end{center}
\end{figure*}

\clearpage

\begin{figure*}
\begin{center}
    \includegraphics[width=110mm,height=100mm]{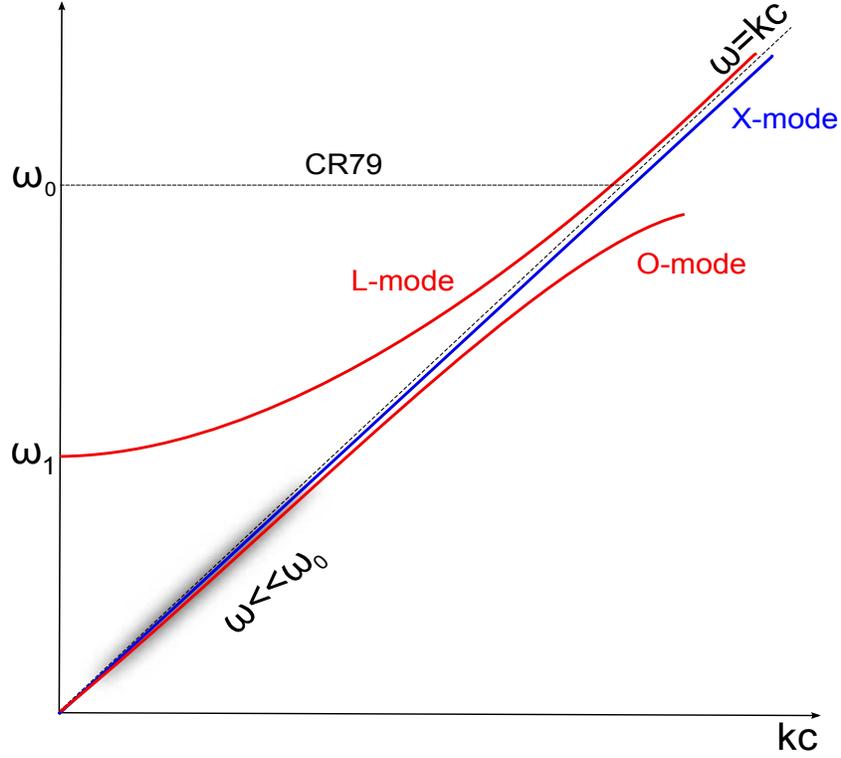}
      \caption{Schematic representation of the electron positron plasma eigen-modes in infinitely strong magnetic
               field in the case of oblique ($\vartheta\neq0$) propagation. The shadowed region represents the
               frequency range $\omega\ll\omega_{\circ}$ characteristic for the soliton curvature radiation, while
               the level $\omega > \omega_{\circ}$ corresponds to the linear Langmuir oscillations used by CR79,
               which, however, cannot emit any coherent curvature radiation. In the case of parallel propagation
               ($\vartheta=0$) the O-mode coincides with the X-mode, while the L-mode becomes a strictly
               longitudinal Langmuir wave. The dispersion curve of the Langmuir waves crosses the $\omega=kc$ line
               at the point $\omega=\omega_{\circ}$.}
    \label{fig2}
\end{center}
\end{figure*}

\clearpage

\begin{figure*}
\begin{center}
    \includegraphics[width=120mm,height=113mm]{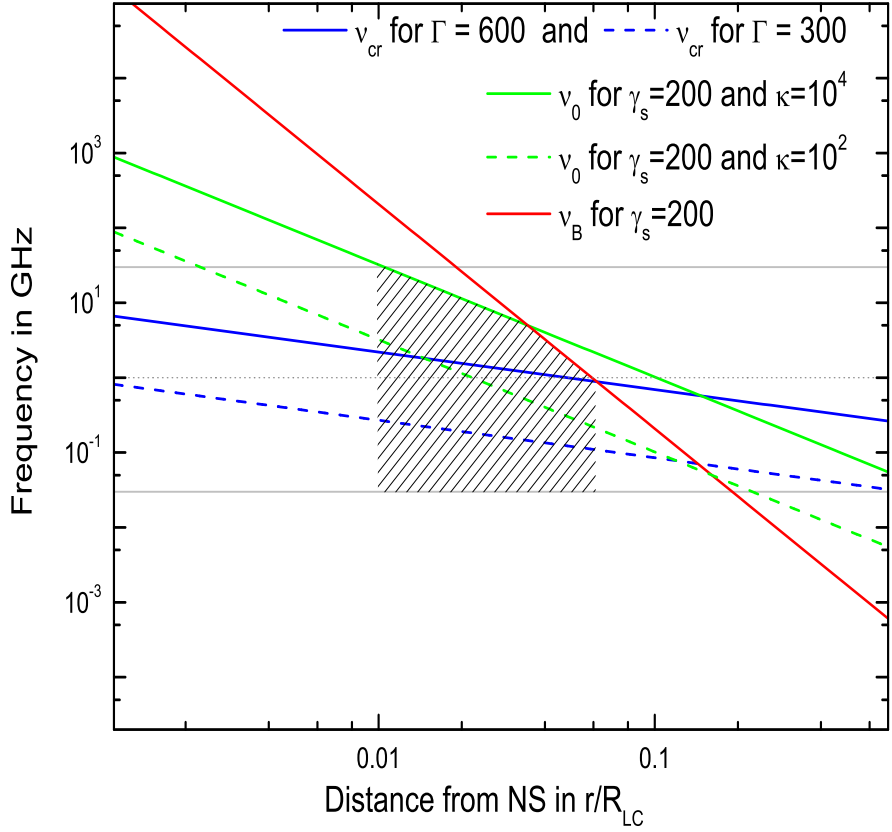}
      \caption{Basic plasma parameters in the observers frame of reference for a typical pulsar with period $P=1$
               sec and ${\dot P}=10^{-15}$. See legend in the figure and the text in Section 3.2 for details. The
               abscissa originates at about 50 km altitude ($r / R_{LC} \sim 10^{-3}$), below which the two stream
               instability is not able to develop. The dashed area represents (from a purely theoretical point of
               view) the most likely region of generation of the observed pulsar radio emission.}
   \label{fig3}
\end{center}
\end{figure*}

\clearpage

\begin{figure*}
\begin{center}
    \includegraphics[width=110mm,height=110mm]{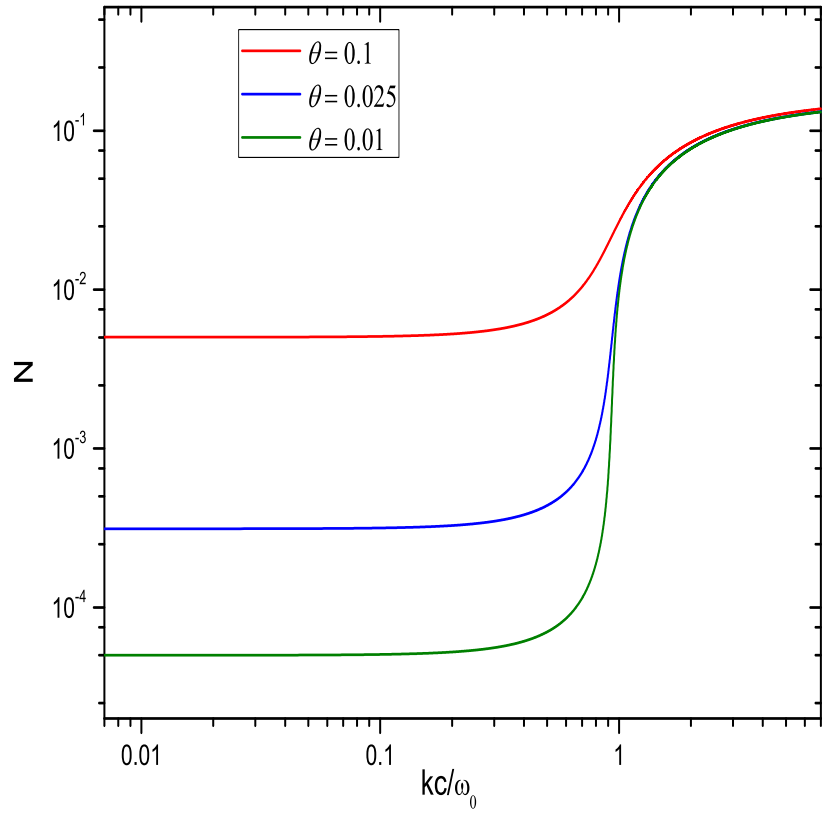}
      \caption{The plot shows the numerical solutions of Equation (\ref{eq_omode}) for three values of
      $\vartheta$. }
   \label{fig4}
\end{center}
\end{figure*}

\end{document}